\pdfoutput=1
\documentclass[12pt]{article}
\usepackage[letterpaper,margin=1in]{geometry}
\usepackage{amsmath,graphicx,booktabs,tabularx,array,setspace,caption,float}
\usepackage[hidelinks]{hyperref}
\usepackage{microtype}
\hypersetup{pdftitle={Revenue Growth and Staff Expense},pdfauthor={Haochuan Wang}}
\begin{document}
\begin{titlepage}
\thispagestyle{empty}
\vspace*{0.10in}
\begin{center}
{\Large\bfseries Revenue Growth and Staff Expense\\[0.12in]
A Public-Company Follow-Up to\\\textit{Eclipse of Rent-Sharing}\par}
\vspace{0.18in}
{\large Haochuan Wang (\href{mailto:hcw@mit.edu}{hcw@mit.edu})\\MIT\par}
\vspace{0.15in}
{\large September 2026\par}
\vspace{0.12in}
\end{center}
\vspace{0.18in}
\begin{center}\textbf{Abstract}\end{center}
\begin{spacing}{1.0}\noindent
Does a widening gap between company revenue and staff expense per employee provide evidence of declining rent-sharing? We examine how reporting selection and sector composition affect that inference, as a follow-up to Acemoglu, He, and le Maire's Eclipse of Rent-Sharing. We assemble 7,000 company-years for 695 firms over 2015--2025. In a fixed sample of 75 staff-expense reporters, nominal revenue per employee rises 83.6\%, while staff expense per employee rises 51.0\%. The pooled expense-to-revenue ratio falls, but the pattern is concentrated in financial firms; among the other 25 reporters, aggregate staff expense per employee grows slightly faster than aggregate revenue per employee. These accounting measures identify neither real productivity nor the allocation of its gains between labor and capital. A separate audit of the original paper's reported estimates finds that some US output effects remain imprecisely estimated, while several Danish estimates support tighter bounds. The fitted reporting-probability model exhibits concentrated weights, substantial two-group imbalance and sectors without reporters; it does not justify generalizing payroll results. We extend the original study's measurement discussion to public-company accounts and qualify a blanket interpretation of no output effects: precision varies across its estimates. Our sector-dependent accounting results neither confirm nor overturn its causal wage and labor-share findings.

\end{spacing}
\vspace{0.18in}
\vfill
\noindent\rule{0.35\textwidth}{0.4pt}\\
{\footnotesize Data: Compustat North America via WRDS, supplied under MIT access.}
\end{titlepage}
\setcounter{page}{1}

\section{The question and our contribution}
A firm can increase revenue while employing fewer people, paying existing workers more, outsourcing work, or raising prices. These possibilities have different implications for workers. Distinguishing them matters when using company growth to interpret evidence about managerial education and rent-sharing.

Acemoglu, He, and le Maire (AHL) report lower wages and labor shares after appointments of business-educated CEOs and interpret the evidence through reduced rent-sharing [1]. Our central question is whether a widening nominal revenue--staff-expense gap supports a similar distributional interpretation in accessible public-company accounts. We assess the sensitivity of that inference to reporting coverage and sector composition. A separate precision audit examines selected reported output estimates in AHL; it does not test the same estimand as our accounting panel.

This paper follows up on Eclipse of Rent-Sharing: The Effects of Managers' Business Education on Wages and the Labor Share in the US and Denmark by Acemoglu, He, and le Maire [1]. Our contribution is an extension and a qualification, rather than a replication of their treatment effects. We extend their distributional question to the measurement of revenue and staff expense in public-company accounts. We qualify a blanket reading of no output effects because selected US estimates do not exclude economically meaningful changes, while several Danish estimates are more precise. Our accounting evidence does not independently confirm or refute their causal wage or labor-share findings: our sample lacks the linked education treatment and worker outcomes needed for that comparison.

\section{Why education, appointment and rent-sharing differ}
Managerial education may affect skills or values. It also correlates with experience, networks and the firms that hire a manager. A board may select a credentialed CEO to implement a restructuring plan already under way. Manager-specific policy differences therefore motivate a selection model, not an assumption that credentials are randomly assigned [2]. Even an identified appointment effect would not show what happens when the same person acquires an MBA.

A simple bargaining relation is w = b + theta*r: worker compensation w equals an outside option b plus a share theta of surplus per worker r. Education, board mandates, incentive contracts and worker voice can affect theta. Technology and demand can affect r. This is an organizing framework; we do not estimate its parameters. Post-appointment contracts may mediate a change or implement a common board mandate. Conditioning on them does not automatically identify that mechanism [3].

For production, labor hours H = Q*h, where Q is quality-adjusted output and h hours per unit. Headcount N = H/average hours. Revenue R = P*Q also contains prices P. Thus higher R/N can reflect prices, demand, hours, outsourcing or organizational boundaries. The task framework distinguishes displacement from output expansion and new-task creation [4]. Without task, price and hours data, our panel cannot separate these channels or identify AI adoption.

\subsection{Joint production and distribution are different questions}
A productivity increase can result from more capital per worker, better skills, organizational change, or technical progress. It should not be attributed entirely to workers or entirely to managers. To organize these channels, consider the illustrative constant-returns production function below. Y denotes real value added, K capital services, H hours worked, q labor quality, and A efficiency. The constant parameter alpha lies between zero and one. This is a theoretical benchmark, not a fitted production model.

\begin{equation}Y=A K^{\alpha}(qH)^{1-\alpha},\qquad 0<\alpha<1.\end{equation}
\begin{equation}\Delta\log(Y/H)=\Delta\log A+\alpha\Delta\log(K/H)+(1-\alpha)\Delta\log q.\end{equation}
Under this constant-alpha specification, real output per hour can grow through efficiency, capital deepening, and labor quality. BLS productivity concepts distinguish output per hour from the joint contribution of inputs [11]. Our extract has neither hours nor capital services or suitable output deflators. Total book assets (at) do not measure the productive flow from capital, and capital expenditure is absent from this extract. Regressing nominal revenue per employee on book assets would not identify this decomposition. Alpha can be interpreted as a factor-income share only under additional competitive-pricing and production assumptions; it is not the CEO compensation share.

Distribution is a separate accounting and bargaining question. BEA defines value added as output less intermediate inputs, with compensation, taxes less subsidies, and gross operating surplus as income components [12]. Revenue in company accounts requires a measurement bridge before it can serve as output, especially for financial firms. Staff expense divided by revenue is therefore not the labor share of value added. Nor is revenue minus staff expense a measure of capital income: it still covers purchased inputs and other costs. A rise in this residual cannot establish that shareholders captured productivity gains.

A CEO is both a worker and, often, an equity holder. Salary and benefits compensate labor; stock and options create exposure to shareholder value. These interests need not coincide with those of all owners or workers. Expanded ExecuComp incentive measures could test whether equity exposure predicts a change in company policy, but CEO compensation does not measure total capital income. Reported staff expense may also include executive pay. We cannot subtract a small, unmatched CEO pilot from company-wide payroll and treat the result as ordinary-worker earnings.

\section{Data and population}
\begin{table}[H]\centering\caption{Data coverage and sample construction}\small\setlength{\tabcolsep}{4pt}\renewcommand{\arraystretch}{1.15}\begin{tabularx}{\linewidth}{>{\hsize=0.7478632478632479\hsize\linewidth=\hsize\raggedright\arraybackslash}X>{\hsize=1.2521367521367521\hsize\linewidth=\hsize\raggedright\arraybackslash}X}\toprule
\textbf{Stage} & \textbf{Observed coverage}\\\midrule
Public historical candidate list & 771 ticker strings; source ends August 18, 2026\\
Raw financial download & 8,298 rows; 696 companies; industrial and financial-services reporting formats\\
Financial analysis sample & 7,000 unique company-years in the industrial reporting format; 695 firms; 2015-2025\\
Fixed positive-outcome sample & 555 firms with employment and revenue in all 11 years\\
CEO records available & 263 CEO-years; 46 CEO IDs; original 24 firms only\\
Validated degree groups & Not observed for the expanded sample\\
\bottomrule\end{tabularx}\end{table}
The financial source is Compustat North America Fundamentals Annual, obtained through WRDS with MIT day-pass access. North America is the database product, not the location of every firm or worker. Outcomes generally describe consolidated operations. In the latest available rows, 653 firms have US incorporation codes and 658 US location codes. These header fields do not establish historical location or CEO background.

The raw extract includes fiscal 2014 and incomplete 2026 records. We restrict outcomes to 2015-2025 and keep financial-services reporting-format rows separate. INDL is a format, not a nonfinancial-industry filter. Historical ticker matching remains provisional: 75 input symbols did not match returned symbols exactly, which is not proof of 75 missing companies. Renames, multiple share classes and reused symbols require dated identifier links.

\section{A direct precision audit of the published estimates}
We recompute normal-approximation intervals from the June 2025 version of AHL [1]. For a log coefficient beta with standard error s, the percentage interval is 100*[exp(beta +/- 1.96*s)-1]. Equivalence within a multiplicative margin m requires rejecting both beta <= log(1-m) and beta >= log(1+m). At 5\% size, this is equivalent to placing the 90\% interval inside those bounds. Margins of 5\% and 10\% are illustrative, not preregistered economic thresholds.

\begin{figure}[H]\centering
\includegraphics[width=\linewidth,height=0.48\textheight,keepaspectratio]{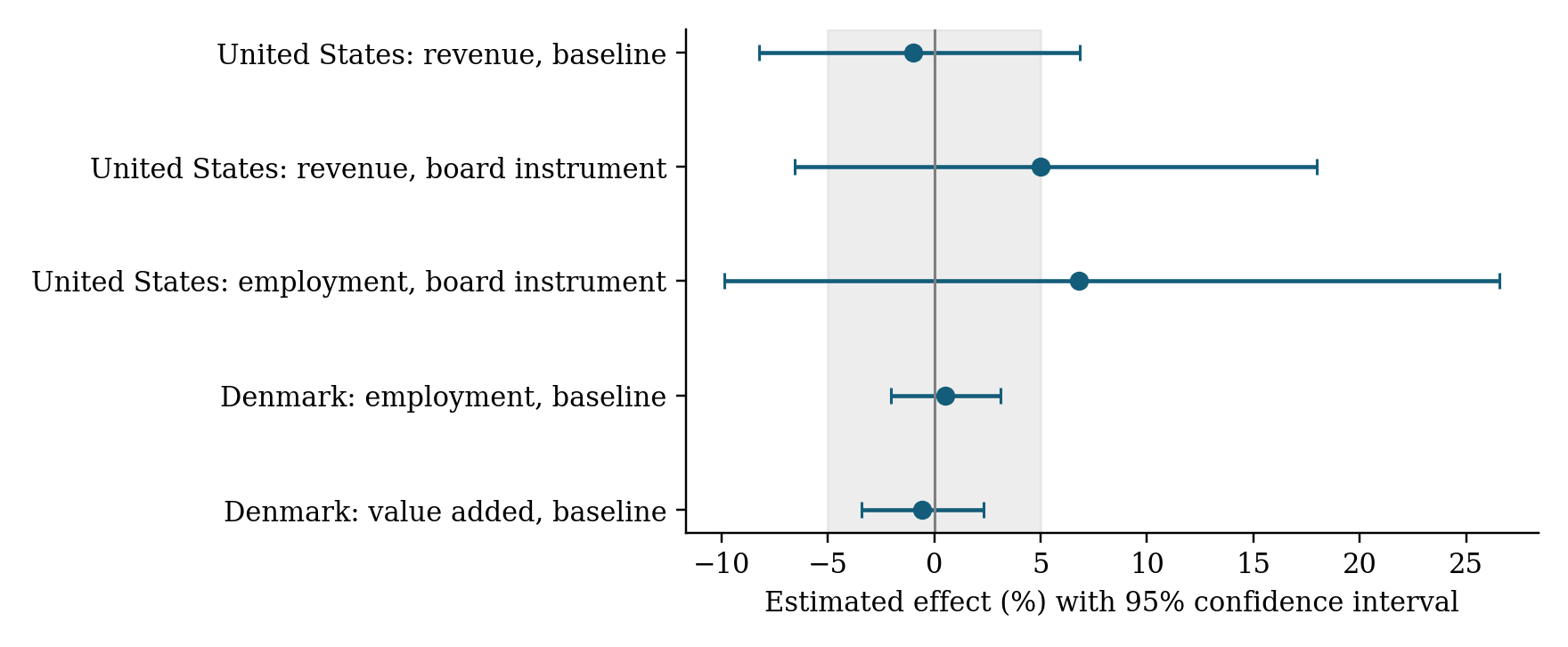}
\caption{Selected reported estimates and 95\% intervals. Shading marks +/-5\%, not an equivalence test based on the 95\% interval. Source: AHL June 2025, Figure 3c, Figure 5d/e and Table 1A; our calculations.}\end{figure}
\begin{table}[H]\centering\caption{Equivalence audit of reported estimates}\small\setlength{\tabcolsep}{4pt}\renewcommand{\arraystretch}{1.15}\begin{tabularx}{\linewidth}{>{\hsize=1.5726495726495726\hsize\linewidth=\hsize\raggedright\arraybackslash}X>{\hsize=0.9401709401709402\hsize\linewidth=\hsize\raggedright\arraybackslash}X>{\hsize=0.7435897435897436\hsize\linewidth=\hsize\raggedright\arraybackslash}X>{\hsize=0.7435897435897436\hsize\linewidth=\hsize\raggedright\arraybackslash}X}\toprule
\textbf{Reported outcome} & \textbf{95\% interval} & \textbf{Equivalence p: +/-5\%} & \textbf{Equivalence p: +/-10\%}\\\midrule
US revenue, baseline & [-8.23, 6.85]\% & .142 & .007\\
US revenue, board IV & [-6.57, 17.98]\% & .499 & .217\\
US employment, board IV & [-9.85, 26.58]\% & .579 & .368\\
Denmark value added & [-3.40, 2.33]\% & .001 & <.001\\
\bottomrule\end{tabularx}\end{table}
The US board-IV revenue estimate, 0.0487 with standard error 0.0595, is consistent with both economically relevant declines and increases. Failure to reject zero is not evidence that meaningful growth effects have been excluded. Conversely, US baseline revenue passes the illustrative +/-10\% equivalence test, and Danish baseline employment and value added pass +/-5\%. It would be incorrect to characterize every null in the paper as uninformative.

This is the note's direct empirical response to AHL: precision varies across outcomes and designs, so claims of little or no output response should be tied to a specific estimand and economically justified margin. We do not establish positive productivity effects. We use reported standard errors, lack covariance matrices for joint or dynamic inference, and do not reproduce the confidential microdata designs.

\section{Employment and revenue in the public-company panel}
Figure 2 holds the set of companies fixed. From 2015 to 2025, aggregate employment rises 14.82\%, nominal revenue 84.69\%, and nominal revenue per employee 60.85\%. These are ratios of sums, not median firm changes. The balanced sample controls changes in which firms are included but selects firms with complete positive outcomes; it is not a representative or causal sample.

\begin{figure}[H]\centering
\includegraphics[width=\linewidth,height=0.48\textheight,keepaspectratio]{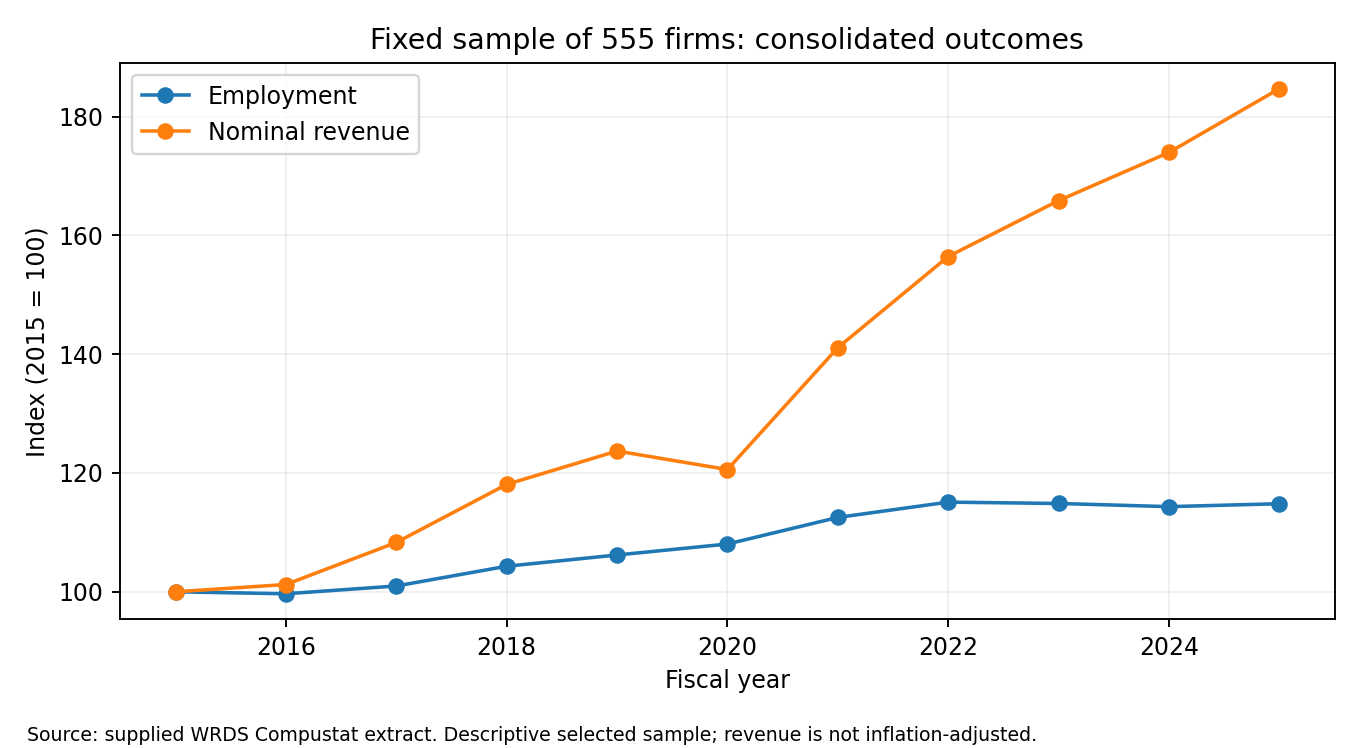}
\caption{Consolidated outcomes for 555 firms with positive employment and revenue in every year. Revenue is nominal. Corporate boundaries can change even for fixed company identifiers. Source: supplied WRDS Compustat extract.}\end{figure}
In the separate 2024-2025 comparison, 605 firms have positive revenue and employment at both endpoints and stable fiscal timing. Median employment growth is 0.00\%; median nominal revenue growth is 5.44\%. Revenue rises while employment falls in 153 firms, or 25.29\%. These are company-level counts, not deductions from aggregate medians.

A narrower 548-firm stable-timing panel shows lower average annual log employment growth in 2023-2025 than in 2016-2019: a difference of -2.69 percentage points, with a firm-bootstrap interval of [-3.70, -1.58]. Nominal revenue growth also slows, by -2.17 points. Appendix A reports all six exploratory contrasts, including inconclusive ones. Calendar periods were selected after viewing earlier trends; no AI treatment is measured.

These patterns illustrate why output and labor-input measures must be separated. They do not contradict AHL: our firms, periods and comparison are different, and no CEO-education treatment is present. Inflation, demand, mergers, spinoffs and outsourcing remain plausible explanations. The descriptive panel cannot by itself adjudicate managerial socialization against a board mandate.

\section{Payroll measurement changes the feasible argument}
Only 1,054 of 7,000 analysis-panel observations report labor and related expense. A 75-firm sample reports payroll with positive employment and revenue throughout 2015-2025. Its pooled payroll/revenue ratio falls, but the sector breakdown matters: 50 firms carry a financial-sector industry classification and only 25 fall in other groups.

\begin{figure}[H]\centering
\includegraphics[width=\linewidth,height=0.48\textheight,keepaspectratio]{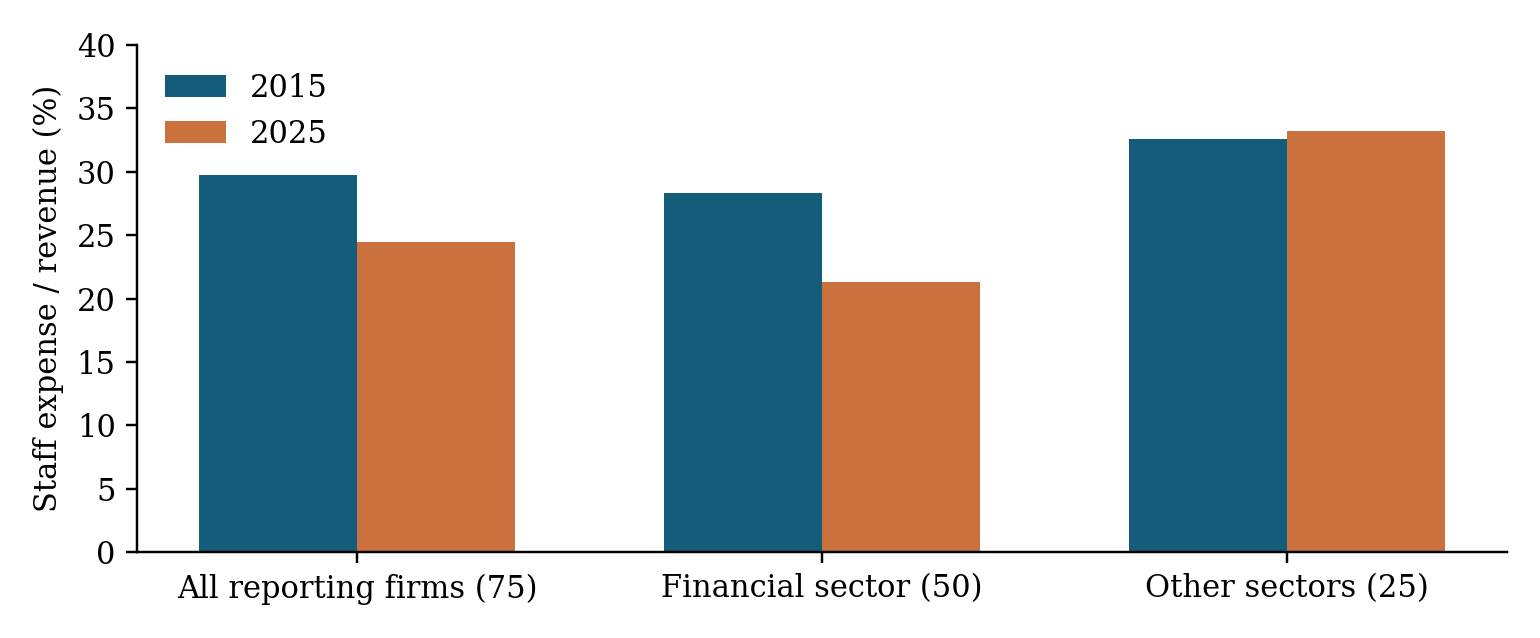}
\caption{Ratios of aggregate payroll to aggregate revenue for consistent reporters. Groups use the industry classification reported for 2025, not verified historical classification. Sector definitions and financial revenue require further validation.}\end{figure}
The financial group's ratio falls from 28.36\% to 21.31\%, whereas the other group's ratio rises from 32.58\% to 33.21\%. Mean firm-specific ratio changes also differ between these groups (Appendix A). Neither statistic is value-added labor share. We therefore reject a broad interpretation that this sample establishes generally declining rent-sharing.

\subsection{A decade of revenue and staff expense per employee}
We now hold the measurement sample fixed across both outcomes. Let R denote nominal revenue (revt), W reported labor and related expense (xlr), and N reported employees (emp). For the same 75 firms with valid observations in every fiscal year from 2015 through 2025, define aggregate ratios r and w. These ratios weight firms implicitly by their employment; they are not averages of individual worker wages or typical-firm growth rates.

\begin{equation}r_t=\frac{\sum_{i\in S}R_{it}}{\sum_{i\in S}N_{it}},\qquad w_t=\frac{\sum_{i\in S}W_{it}}{\sum_{i\in S}N_{it}},\qquad s_t=\frac{\sum_{i\in S}W_{it}}{\sum_{i\in S}R_{it}}.\end{equation}
\begin{equation}\Delta\log s_t=\Delta\log w_t-\Delta\log r_t.\end{equation}
This identity separates the observed compensation-cost and revenue measures without assigning a causal channel. A falling s means that staff expense per employee grew more slowly than revenue per employee in these accounting totals. It does not establish a fall in real wages, a change in the compensation of a continuing worker, or the fraction of real productivity growth paid to labor. Workforce composition, working hours, nonwage benefits and corporate boundaries may all change.

\begin{figure}[H]\centering
\includegraphics[width=\linewidth,height=0.48\textheight,keepaspectratio]{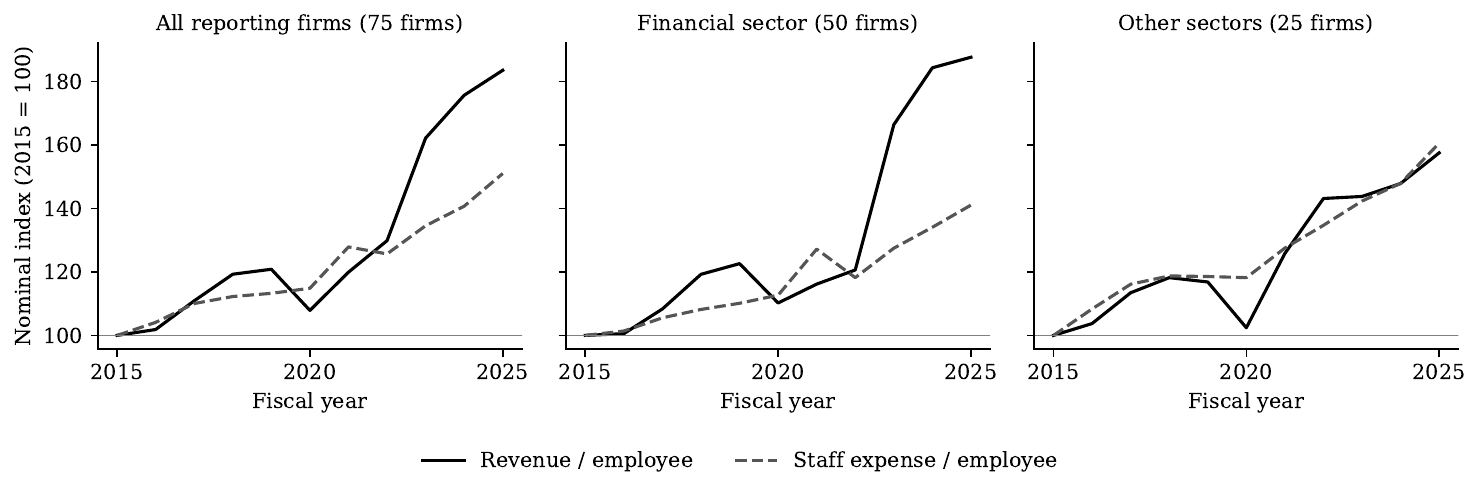}
\caption{Nominal revenue and staff expense per employee, fixed 2015--2025 reporting samples. Both series use the same firms and employment denominator within each panel. Industry groups use the classification reported for 2025. Source: supplied WRDS Compustat extract; new accounting decomposition.}\end{figure}
\begin{table}[H]\centering\caption{Nominal per-employee changes, 2015--2025}\small\setlength{\tabcolsep}{4pt}\renewcommand{\arraystretch}{1.15}\begin{tabularx}{\linewidth}{>{\hsize=1.0683760683760684\hsize\linewidth=\hsize\raggedright\arraybackslash}X>{\hsize=0.42735042735042733\hsize\linewidth=\hsize\raggedright\arraybackslash}X>{\hsize=1.0683760683760684\hsize\linewidth=\hsize\raggedright\arraybackslash}X>{\hsize=1.0683760683760684\hsize\linewidth=\hsize\raggedright\arraybackslash}X>{\hsize=1.3675213675213675\hsize\linewidth=\hsize\raggedright\arraybackslash}X}\toprule
\textbf{Sample} & \textbf{Firms} & \textbf{Revenue / employee growth} & \textbf{Staff expense / employee growth} & \textbf{Staff expense / revenue, 2015 to 2025}\\\midrule
All reporters & 75 & 83.56\% & 51.03\% & 29.75\% to 24.47\%\\
Financial sector & 50 & 87.67\% & 41.06\% & 28.36\% to 21.31\%\\
Other sectors & 25 & 57.53\% & 60.59\% & 32.58\% to 33.21\%\\
\bottomrule\end{tabularx}\end{table}
The pooled sample exhibits a substantial nominal divergence: revenue per employee rises 83.56\%, compared with 51.03\% for staff expense per employee. The staff-expense/revenue ratio falls 5.27 percentage points. However, finance accounts for 50 of the 75 firms. The other 25 show staff expense per employee growing slightly faster than revenue per employee, and a 0.63-point increase in the expense ratio. The pooled pattern cannot be generalized into a universal decline in sharing. Header-based sector classification and financial revenue definitions need further audit.

These growth rates differ from the 60.85\% revenue-per-employee increase in the 555-firm sample because the samples differ. Comparing that larger-sample rate with the 75-firm staff-expense rate would mix populations. The new figure compares like coverage with like coverage; it does not make the reporters representative of all firms.

Across the 75 firms, additional staff expense equals 19.23\% of additional nominal revenue between the endpoints; the corresponding figures are 15.42\% in finance and 34.24\% elsewhere. These are ratios of changes in aggregate accounting amounts, not estimates of a marginal propensity to share an exogenous rent shock. The remaining 80.77\% in the pooled calculation is not a capital share. We retain this calculation in the underlying CSV for transparency but do not use it as a productivity-allocation result.

The endpoint figures are descriptive finite-sample totals, with no null-hypothesis test attached. Confidence intervals for a hypothetical population would not repair selection or measurement error. An exact log-identity check, unique firm-year check, and fixed-sample check accompany the saved output. Changes in consolidation, fiscal timing and staff-expense definitions remain alternative explanations for persistent within-firm accounting changes. The following influence checks assess aggregation, not those measurement mechanisms.

\subsection{Within-firm changes, aggregation and influence}
Keeping the same firms does not keep their economic weights fixed. Write the aggregate expense ratio as the revenue-weighted sum of firm ratios. The midpoint decomposition below is an exact accounting identity, allocating the change to movements in company ratios and movements in revenue weights. It is descriptive and does not assign a causal mechanism.

\begin{equation}\Delta s=\sum_i\bar a_i\Delta s_i+\sum_i\bar s_i\Delta a_i,\qquad a_{it}=R_{it}/\sum_jR_{jt}.\end{equation}
Bars denote the average of 2015 and 2025 endpoint values. For all 75 reporters, the aggregate ratio declines 5.272 percentage points: the within-company ratio component is -5.291 points and the revenue-weight component is +0.020 points. The decline therefore is not an artifact of changing company revenue weights alone. Neither component measures a change in value-added labor share.

\begin{table}[H]\centering\caption{Aggregation and single-company influence checks}\small\setlength{\tabcolsep}{4pt}\renewcommand{\arraystretch}{1.15}\begin{tabularx}{\linewidth}{>{\hsize=0.9081196581196581\hsize\linewidth=\hsize\raggedright\arraybackslash}X>{\hsize=1.014957264957265\hsize\linewidth=\hsize\raggedright\arraybackslash}X>{\hsize=1.014957264957265\hsize\linewidth=\hsize\raggedright\arraybackslash}X>{\hsize=0.9081196581196581\hsize\linewidth=\hsize\raggedright\arraybackslash}X>{\hsize=1.1538461538461537\hsize\linewidth=\hsize\raggedright\arraybackslash}X}\toprule
\textbf{Group} & \textbf{Aggregate change (pp)} & \textbf{Equal-firm mean (pp)} & \textbf{Median firm (pp)} & \textbf{Leave-one-out range (pp)}\\\midrule
All 75 & -5.272 & -3.338 & -3.362 & [-5.965, -4.635]\\
Finance 50 & -7.043 & -4.915 & -4.622 & [-8.213, -6.276]\\
Other 25 & 0.633 & -0.183 & -0.291 & [0.017, 1.392]\\
\bottomrule\end{tabularx}\end{table}
The pooled and finance-group aggregate declines retain their signs after each single-company omission. The other-sector aggregate increase also retains its sign, but the equal-firm mean and median changes are slightly negative. Consequently, the statement that staff expense keeps pace with revenue outside finance applies to aggregate dollars, not to the typical company. The within-ratio components are -7.722 points for finance and +0.558 points for the other group; changing revenue weights contribute +0.679 and +0.075 points respectively. These checks clarify which aggregation supports each claim, while leaving financial accounting definitions and reporting selection unresolved.

\subsection{Can reporting weights repair selection?}
We estimate a payroll-reporting probability in fiscal 2015 among 631 firms with positive employment, revenue and assets; 95 report payroll. The exploratory logistic model uses log employment, revenue and assets and broad industry classifications. Inverse-observation weighting targets the mean firm payroll/revenue ratio only if conditional observation-at-random, positivity and model adequacy hold. It is not treatment weighting for MBA appointments [5].

\begin{table}[H]\centering\caption{Payroll-reporting weighting diagnostics}\small\setlength{\tabcolsep}{4pt}\renewcommand{\arraystretch}{1.15}\begin{tabularx}{\linewidth}{>{\hsize=1.1965811965811965\hsize\linewidth=\hsize\raggedright\arraybackslash}X>{\hsize=0.8034188034188035\hsize\linewidth=\hsize\raggedright\arraybackslash}X}\toprule
\textbf{Diagnostic} & \textbf{Observed result}\\\midrule
Largest absolute standardized difference & 1.43 before; 0.78 after weighting\\
Reporter effective sample size & 18.23 from 95 observed reporters\\
Largest normalized reporter weight & 18.28\% of reporter weight\\
Reporting cells & 7 sectors without reporters; 1 with complete reporting\\
\bottomrule\end{tabularx}\end{table}
This fitted weighting specification retains substantial two-group imbalance, has concentrated weights, and encounters seven sectors without reporters. We retain its numerical output and bootstrap draws for transparency but do not use the adjusted ratio as representative evidence. Detailed diagnostic output is retained in the research archive. This failed attempt is substantively informative: accessible payroll data do not support an unrestricted public-company rent-sharing study.

\subsection{Estimands and assumptions for inverse-probability weighting}
The completed weighting exercise concerns whether payroll is observed. Let S\_i equal one if payroll is reported in 2015, Z\_i=W\_i/R\_i, and X\_i include the observed baseline firm characteristics. The target is the equal-firm mean of Z across the 631 eligible baseline firms. It is not the aggregate payroll/revenue ratio used in the decade analysis. Defining p(X)=Pr(S=1|X), the normalized inverse-observation estimator is:

\begin{equation}\widehat\mu_{\mathrm{H}}=\frac{\sum_i S_i Z_i/\widehat p(X_i)}{\sum_i S_i/\widehat p(X_i)}.\end{equation}
Its justification requires three substantive conditions. First, conditional mean exchangeability of reporting: E[Z|X,S=1]=E[Z|X]. After conditioning on the included characteristics, reporters must be informative about the missing payroll ratios. This can fail if firms with unusually high labor costs disclose them differently. Second, positivity: every target covariate profile must have a nonzero reporting probability. A fitted logistic score above zero does not supply observations in an empty sector. Third, the probability model must adequately approximate the reporting process. Normalization stabilizes the scale but does not fix misspecification or lack of support. The outcome and reporting definitions must also be consistently measured.

The implementation uses an L2-penalized logistic regression with C=1, standardized log employment, revenue and assets, and two-digit NAICS indicators. This regularized, low-dimensional choice is a diagnostic specification rather than a claim that the logistic form is true. Contemporaneous 2015 variables are used to model 2015 reporting, not to adjust a CEO treatment. Historical header classification remains a limitation. An observation-IPW identification assumption cannot be reused as an assumption about random CEO assignment.

The internal balance calculation compares reporters and nonreporters before and after weighting. For balance diagnostics, reporters receive 1/p and nonreporters 1/(1-p). In contrast, the outcome estimator uses only reporting firms because Z is missing for others. The standardized mean difference for each covariate uses its original pooled within-group standard deviation, held fixed before and after weighting. This prevents apparent improvement caused solely by changing the denominator. It remains a two-group diagnostic: the direct balance check for the target mean compares weighted reporters with the full eligible target sample, using target-standard-deviation scaling. This two-group diagnostic is distinct from that direct target-balance test. Indicator variables use the same convention; the 0.1 reference line is a heuristic rather than a hypothesis-test cutoff [5].

\begin{equation}\mathrm{SMD}_j=\frac{\bar X_{1j,w}-\bar X_{0j,w}}{\sqrt{(s^2_{1j}+s^2_{0j})/2}},\qquad \mathrm{ESS}=\frac{(\sum_i w_i)^2}{\sum_i w_i^2}.\end{equation}
Balance is checked without choosing the model to obtain an attractive payroll effect. Small standardized differences would support comparability on these measured variables only; they would not establish the reporting assumption. Here the maximum absolute reporter-versus-nonreporter imbalance remains 0.78 and the reporting effective sample size falls to 18.23. Seven sectors contain no reporters, limiting identification of their missing outcomes. A sector containing only reporters does not violate positivity for an observation-weighted mean; the upper bound p<1 is unnecessary for that estimand. These diagnostics do not justify a full-target interpretation of this fitted estimator. Failure of this specification is not proof that every observation-weighting specification would fail. Weight capping reduces the influence of extreme observations but changes the estimator and can introduce bias. It cannot reconstruct unobserved sector outcomes.

Inverse probability weighting addresses observed selection, not heterogeneous error variance by itself. CEOs may differ in unmeasured ability, background and mandates even at the same propensity score. Robust or clustered inference addresses some sampling uncertainty; it does not remove those confounders. The existing 499 bootstrap repetitions refit the standardization and propensity model in each firm resample. This propagates fitted-weight uncertainty for the chosen procedure, conditional on its assumptions, but does not make the failed diagnostic a valid population estimate.

\section{CEO appointment effects and identification}
Freeze eligibility and firm/board covariates before the succession announcement. Compare within pre-event industry and calendar cohorts, balancing earlier firm size, growth, distress and supported executive characteristics. Age, birth cohort and calendar year cannot all have unrestricted separate effects. Birthplace, nationality and migration status must not be inferred from names or company location. CEO fixed effects absorb fixed credentials and do not recover their causal effect.

\subsection{CEO succession: estimand and identification}
The proposed unit is a firm succession event. At time zero, an eligible firm replaces a verified nonbusiness incumbent with a successor whose education was completed before appointment. D=1 denotes a business-educated successor and D=0 a verified nonbusiness successor. Mixed or unknown credentials require separate coding; an MBA indicator is narrower than AHL business education. Primary outcomes would be log employment and log nominal revenue at one and three fiscal years after appointment. Reported staff expense is secondary and restricted to a documented reporting population. These choices should be fixed before analyzing the expanded education links.

\begin{equation}\tau_{g,h}=E[Y_{i,g+h}(1)-Y_{i,g+h}(0)\mid D_i=1,G_i=g].\end{equation}
This cohort-and-horizon ATT asks what happened after the initial appointment relative to the counterfactual successor type for treated firms. It is not the effect of giving an MBA to the same person. Use contemporaneous succession controls in the same pre-event industry and calendar cohort and require common support on pre-announcement covariates. Avoid selecting firms for remaining in the index or retaining the same CEO through the outcome horizon. For an initial-assignment estimand, later turnover is part of the realized path; conditioning on retention would select on a post-treatment variable. Repeated or overlapping events need a predeclared rule, such as the first eligible event per firm in the analysis period.

Pre-event covariates should include firm scale, revenue and employment growth histories, financial condition, board composition and existing compensation policy. Supported successor age, prior experience, education timing and background can improve comparability but may be components of a joint hiring choice. Adjusting for them changes the substantive comparison and cannot isolate the schooling intervention. Industry composition is addressed through within-industry comparisons and historical industry-by-calendar controls, not by assuming an MBA indicator absorbs technology or finance differences. Neither immigrant status nor birthplace can be inferred from names.

\begin{equation}\begin{split}E[Y_{g+h}(0)-Y_{g-1}(0)\mid D=1,X,G=g]\\=E[Y_{g+h}(0)-Y_{g-1}(0)\mid D=0,X,G=g].\end{split}\end{equation}
The conditional parallel-trends assumption above is the central restriction for a difference-in-differences interpretation. It allows different baseline outcome levels but requires comparable counterfactual changes after conditioning. Additional requirements are no anticipation before the selected baseline, consistent treatment and outcome definitions, overlap, and no material spillovers into comparison firms. A restructuring mandate known before the appointment can violate parallel trends and no anticipation even if measured covariates balance. Event-time leads and sensitivity analysis can expose incompatibilities but cannot prove the assumption [9].

For an ATT comparison, treatment-propensity weights are one for treated events and e(X)/(1-e(X)) for controls, normalized within the relevant cohort. A doubly robust difference-in-differences procedure combines propensity weighting and a model for untreated outcome changes [6,7]. Under the identification assumptions, correct specification of either nuisance component can suffice; neither correct component rescues a false parallel-trends assumption. We have not computed these weights or an MBA treatment effect because validated expanded CEO education and succession dates are unavailable.

The treatment is assigned at a succession, while outcomes repeat within firms. Inference should therefore keep the entire firm trajectory together and refit nuisance estimates on each bootstrap draw. Seven thousand firm-years do not represent seven thousand independent appointments. The number of treated events, control events, firms and effective weighted observations must all be reported. With only eleven fiscal years, firm-only resampling conditions on the observed macroeconomic history and may miss common shocks. Two-way or calendar-cohort sensitivity checks require care with so few time clusters; additional bootstrap draws cannot solve a shortage of independent clusters.

Compensation contracts and board mandates provide competing explanations for an appointment association. A pre-existing equity incentive may be a confounder; a newly negotiated option grant may be a mediator. Adding post-appointment grants to a regression does not prove mediation and can induce bias. A mechanism study needs its own temporal ordering and mediator-outcome assumptions [3]. The present contribution is an explicit design and measurement audit; the board-regime explanation remains a hypothesis, not a demonstrated alternative model.

\subsection{Selection and external validity}
The 695-company panel includes 79 firms marked inactive in their latest available records, but this does not remove survivorship. Future S\&P entry, historical ticker recovery, positive adjacent outcomes and all-year reporting each select observations. Acquisitions can remove a company without eliminating its jobs; missing data cannot be set to zero. The causal sample should be defined before appointment, remain unbalanced, and include a dated exit and identifier audit. Observation weights and selection bounds need explicit assumptions [10].

\section{Conclusion}
AHL's wage and rent-sharing evidence warrants a follow-up that separates the distributional question from what public-company accounts can identify. Our findings extend Eclipse of Rent-Sharing through a sector and reporting-selection analysis, and qualify a blanket interpretation of no output response through the precision audit. The qualification concerns the strength of selected output conclusions, not evidence that output increased. We neither confirm nor overturn the original causal wage and labor-share results. The present sample does not identify CEO-education effects, and staff expense divided by revenue is not labor's share of value added.

The new decade comparison sharpens the distribution question. The same 75 firms show nominal revenue per employee growing 83.6\% and staff expense per employee growing 51.0\%. Yet the finance and other-sector results differ in direction for the expense ratio. That heterogeneity makes measurement and reporting selection central to any proposed rent-sharing interpretation. Neither this divergence nor the 19.2\% incremental payroll-to-revenue ratio assigns productivity gains to shareholders.

The within-firm decomposition and single-company omissions show that the pooled decline is not driven solely by changing revenue weights or one company. The nonfinance aggregation contrast nevertheless qualifies statements about typical firms. Consistent expense definitions and major acquisitions require separate accounting validation. Sector-specific output deflators and hours would support a more credible productivity proxy; purchased-input costs and an accounting bridge to value added are needed for labor shares. A national price index alone would not turn multinational consolidated revenue into a quantity measure. Product mix and changes in company boundaries remain relevant even after inflation adjustment.

A complementary mechanism exercise would link dated CEO appointments, education and pre-appointment equity incentives across the expanded firms. Pre-existing incentives can help describe the board regime; post-appointment changes require a separate mediation design. School and graduation-cohort comparisons need adequate overlap within industries and appointment cohorts. Such tests should be organized around a small declared family of hypotheses, with the full set of estimates and intervals retained, rather than selecting only statistically significant results.

An AI extension requires observed deployment dates or task exposure and a credible comparison group. A post-2022 calendar indicator cannot separate AI from demand, interest rates or pandemic recovery. Hiring, separations, hours, occupations and retraining outcomes would distinguish displacement from task reallocation more effectively than total headcount alone. Until those inputs exist, the paper supports a transparent measurement agenda and a qualified interpretation of selected output estimates, while leaving the causal distribution and transition-policy questions open.

\clearpage\begingroup\small\setstretch{1.0}\section*{References}
\noindent [1] Acemoglu, D., A. X. He, and D. le Maire (2025). Eclipse of Rent-Sharing: The Effects of Managers' Business Education on Wages and the Labor Share in the US and Denmark. June 2025 manuscript; AER listed forthcoming when checked September 13, 2026. Quantitative audit uses this manuscript version. \href{https://lemaire.dk/EclipseOfRentSharing2025June.pdf}{Source}.\par
\noindent [2] Bertrand, M., and A. Schoar (2003). Managing with Style: The Effect of Managers on Firm Policies. Quarterly Journal of Economics 118(4), 1169-1208. \href{https://doi.org/10.1162/003355303322552775}{Source}.\par
\noindent [3] Imai, K., L. Keele, and T. Yamamoto (2010). Identification, Inference and Sensitivity Analysis for Causal Mediation Effects. Statistical Science 25(1), 51-71. \href{https://arxiv.org/abs/1011.1079}{Source}.\par
\noindent [4] Acemoglu, D., and P. Restrepo (2019). Automation and New Tasks: How Technology Displaces and Reinstates Labor. Journal of Economic Perspectives 33(2), 3-30. \href{https://www.nber.org/papers/w25684}{Source}.\par
\noindent [5] Austin, P. C., and E. A. Stuart (2015). Moving towards best practice when using inverse probability of treatment weighting using the propensity score to estimate causal treatment effects in observational studies. Statistics in Medicine 34, 3661-3679. \href{https://pmc.ncbi.nlm.nih.gov/articles/4626409/}{Source}.\par
\noindent [6] Sant'Anna, P. H. C., and J. Zhao (2020). Doubly Robust Difference-in-Differences Estimators. Journal of Econometrics 219(1), 101-122. \href{https://arxiv.org/abs/1812.01723}{Source}.\par
\noindent [7] Callaway, B., and P. H. C. Sant'Anna (2021). Difference-in-Differences with Multiple Time Periods. Journal of Econometrics 225(2), 200-230. \href{https://arxiv.org/abs/1803.09015}{Source}.\par
\noindent [8] Abadie, A., S. Athey, G. W. Imbens, and J. M. Wooldridge (2023). When Should You Adjust Standard Errors for Clustering? Quarterly Journal of Economics 138(1), 1-35. \href{https://www.nber.org/papers/w24003}{Source}.\par
\noindent [9] Rambachan, A., and J. Roth (2023). A More Credible Approach to Parallel Trends. Review of Economic Studies 90(5), 2555-2591. \href{https://www.jonathandroth.com/assets/files/HonestParallelTrends_Main.pdf}{Source}.\par
\noindent [10] Lee, D. S. (2009). Training, Wages, and Sample Selection: Estimating Sharp Bounds on Treatment Effects. Review of Economic Studies 76(3), 1071-1102. \href{https://www.nber.org/papers/w11721}{Source}.\par
\noindent [11] U.S. Bureau of Labor Statistics. Handbook of Methods: Productivity Measures, Business Sector and Major Subsectors, Concepts. Accessed September 2026. \href{https://www.bls.gov/opub/hom/msp/concepts.htm}{Source}.\par
\noindent [12] U.S. Bureau of Economic Analysis. What is industry value added? Accessed September 2026. \href{https://www.bea.gov/help/faq/184}{Source}.\par
Data sources: S\&P Global Compustat North America Fundamentals Annual via WRDS/MIT day pass, user export received September 13, 2026; public historical candidate symbols from github.com/fja05680/sp500 and current symbols from github.com/datasets/s-and-p-500-companies. These secondary symbol lists are not verified CRSP membership. The supplied IPW lecture informed the implementation; its author/date were not identified. Licensed microdata are not redistributed with this article.

\endgroup\clearpage\appendix\small\setstretch{1.03}
\section{Exploratory tests and uncertainty}
The following is the complete six-test family, not a selection of significant results. A written batch protocol preceded execution but followed earlier descriptive analysis. All estimates remain exploratory. We used 2,999 firm resamples, seed 20260913; percentile intervals and recentered bootstrap-null p-values. Holm adjustment covers the six p-values; intervals are pointwise and unadjusted.

\begin{table}[H]\centering\caption{Complete exploratory test family}\small\setlength{\tabcolsep}{4pt}\renewcommand{\arraystretch}{1.15}\begin{tabularx}{\linewidth}{>{\hsize=2.0085470085470085\hsize\linewidth=\hsize\raggedright\arraybackslash}X>{\hsize=0.37393162393162394\hsize\linewidth=\hsize\raggedright\arraybackslash}X>{\hsize=0.5876068376068376\hsize\linewidth=\hsize\raggedright\arraybackslash}X>{\hsize=1.3888888888888888\hsize\linewidth=\hsize\raggedright\arraybackslash}X>{\hsize=0.6410256410256411\hsize\linewidth=\hsize\raggedright\arraybackslash}X}\toprule
\textbf{Contrast} & \textbf{Firms} & \textbf{Estimate} & \textbf{95\% interval} & \textbf{Holm p}\\\midrule
Employment slowdown & 548 & -2.69 & [-3.70, -1.58] & .002\\
Nominal revenue slowdown & 548 & -2.17 & [-3.19, -1.09] & .002\\
Nonfinance employment slowdown & 468 & -2.72 & [-3.85, -1.53] & .002\\
Earlier employment period contrast & 548 & 0.90 & [-0.37, 2.18] & .343\\
Payroll change: finance minus other & 75 & -4.73 & [-8.21, -1.21] & .026\\
Slowdown: large minus small & 274 & 0.93 & [-2.56, 4.42] & .604\\
\bottomrule\end{tabularx}\end{table}
Growth contrasts compare mean annual 100*log(current/prior) in 2023-2025 versus 2016-2019. The earlier-period contrast compares 2018-2019 with 2016-2017. Payroll is a difference between group mean firm-specific ratio changes, in percentage points. Large and small refer to top/bottom quartiles of 2015 employment; this is not a measure of firm age. The reported industry classification is unverified historically. The simulation p-value floor is 1/3000, not zero.

Firm resampling assumes independent clusters from a hypothetical selected-firm population, conditional on realized calendar shocks. It does not capture common macroeconomic uncertainty, historical selection or unobserved confounding. Seven of the 555 fixed-outcome firms fail the all-year stable-timing screen, leaving 548. Balanced samples remain selected on reporting. The earlier-period and size contrasts are inconclusive; their wide intervals do not establish no effect.

\begin{table}[H]\centering\caption{Recent-pair descriptive bootstrap intervals}\small\setlength{\tabcolsep}{4pt}\renewcommand{\arraystretch}{1.15}\begin{tabularx}{\linewidth}{>{\hsize=1.5705128205128205\hsize\linewidth=\hsize\raggedright\arraybackslash}X>{\hsize=0.4807692307692308\hsize\linewidth=\hsize\raggedright\arraybackslash}X>{\hsize=0.9487179487179487\hsize\linewidth=\hsize\raggedright\arraybackslash}X}\toprule
\textbf{2025 descriptive statistic} & \textbf{Point} & \textbf{95\% bootstrap interval}\\\midrule
Median employment growth & 0.00\% & [0.00\%, 0.53\%]\\
Median nominal revenue growth & 5.44\% & [4.84\%, 6.40\%]\\
Revenue up / employment down & 25.29\% & [21.82\%, 28.76\%]\\
\bottomrule\end{tabularx}\end{table}
These intervals resample the 605 recent firm pairs. Rounding and a mass of zero employment changes make the median interval discrete; it is not an equivalence test. Data screens retain rather than winsorize large changes: 190 firm-year changes in the broader adjacent-year exercise are flagged for corporate-action review. Raw data, inclusion rules, weights, seed and bootstrap draws are retained in the project.

\end{document}